\documentclass[12pt]{article}%
\usepackage{graphicx}%
\usepackage[normalem]{ulem}
\usepackage{amsmath,amsthm,amsfonts,
amsbsy,amssymb,upref,enumerate,bigstrut,
color,mathtools,mathrsfs,float,bm,dsfont,scalefnt,mathtools}
\usepackage{natbib}
\usepackage{lipsum}
\usepackage{lmodern}
\usepackage{stmaryrd}
\usepackage{authblk} 
\usepackage{subfigure}
\usepackage{pgfplots}
\usetikzlibrary{pgfplots.fillbetween}
\usepackage{longtable}

\usepackage{amsmath}
\usepackage{amssymb}

\usepackage{algorithm}
\usepackage{algorithmic}   

\usepackage{booktabs}

\usepackage{caption}
\usepackage{tikz}
\usetikzlibrary{intersections}

\usepackage[colorlinks=true,linkcolor=blue,citecolor=blue]{hyperref}

\usepackage[left=1in,top=1.2in,right=0.7in]{geometry}

\usepackage{booktabs}

\numberwithin{equation}{section} 

\makeatletter
\def\@seccntformat#1{\@ifundefined{#1@cntformat}%
	{\csname the#1\endcsname\quad}
	{\csname #1@cntformat\endcsname}
}
\makeatother

\newif\ifShowComments
\ShowCommentstrue
\def\strutdepth{\dp\strutbox}
\def\druk#1{\strut\vadjust{\kern-\strutdepth
        {\vtop to \strutdepth{%
                \baselineskip\strutdepth\vss
                        \llap{\hbox{#1}\quad}\null}}}}

\title{\bf
Quantile autoregressive moving average models for ratio-based bounded time series
}

\author[1,2]{Helton Saulo\thanks{Corresponding author: heltonsaulo@gmail.com}}
\author[1]{Roberto Vila\thanks{rovig161@gmail.com}}
\author[3]{Filidor Vilca\thanks{fily@unicamp.br}}
\affil[1]{Department of Statistics, University of
	 Bras\'ilia, Bras\'ilia, Brazil}
\affil[2]{
Department of Economics, Federal University of Pelotas, Pelotas, Brazil}
\affil[2]{
Department of Statistics,  State University of Campinas, Campinas, Brazil}

\begin{document}
\maketitle

\begin{abstract}
This paper proposes the quantile unit-log-symmetric autoregressive moving average (QULS--ARMA) model for bounded time series on the open unit interval $(0,1)$. The model extends the unit-log-symmetric family by introducing a quantile-based reparameterization and embedding autoregressive and moving-average dynamics directly in the conditional quantile, thereby overcoming limitations of mean-based approaches and providing a coherent framework for proportion data arising from ratios of dependent positive variables. The proposed specification accommodates asymmetric behavior and heavy tails through flexible log-symmetric kernels, including the normal and Student-$t$ distributions. Parameter estimation is carried out via conditional maximum likelihood, and asymptotic properties are established. Monte Carlo simulations and an empirical application to hydroelectric energy storage proportions in Brazil assess the finite-sample performance and practical advantages of the QULS--ARMA model. The results show the good performance of the proposed estimators across a range of scenarios and kernel specifications.
\end{abstract}
\smallskip
\noindent
{\small {\bfseries Keywords.} {Unit-log-symmetric distribution $\cdot$ ARMA models $\cdot$ Bounded distributions $\cdot$ Monte Carlo simulation  $\cdot$ Proportion data }}
\\
{\small{\bfseries Mathematics Subject Classification (2010).} {MSC 60E05 $\cdot$ MSC 62Exx $\cdot$ MSC 62Fxx.}}


\section{Introduction}
\noindent

Bounded time series data, such as rates and proportions, commonly arise in a wide
range of applications in economics, finance, biology, and environmental sciences.
Typical examples include market shares, default probabilities, employment ratios,
and energy use fractions. Modeling such data is challenging because standard time
series models defined on the real line may yield predictions outside the unit interval
$(0,1)$. As a result, several bounded autoregressive moving average (ARMA) frameworks
have been developed in recent years, most notably those based on the beta distribution
\citep{FerrariCribari2004, RochaCribari2009} and the Kumaraswamy distribution
\citep{Kuma1980,Bayer2017}, which typically model the conditional mean.
More recently, \citet{Ribeiroeal2023} proposed a unit Burr XII quantile autoregressive
moving average (UBXII--ARMA) model, extending static unit distributions to dynamic
frameworks through ARMA-type dynamics on conditional quantiles.

Despite their flexibility, those existing methodologies for bounded time series
present important conceptual limitations when applied to asymmetric proportion data. In particular, the beta and Kumaraswamy ARMA-type models
are formulated in terms of the conditional mean, which may be an inadequate central tendency measure when the distribution is skewed. Quantile-based approaches partially address this issue by modeling the conditional
quantiles directly. A notable example is the UBXII--ARMA model proposed by
\citet{Ribeiroeal2023}, which introduces ARMA-type dynamics in the conditional quantile
of a unit Burr XII distribution. Although this framework represents a substantial
advance over mean-based models, its probabilistic construction is not explicitly
linked to a ratio-type data-generating mechanism. As a consequence, the underlying
stochastic structure does not directly reflect situations in which the observed
proportion arises from dependent random variables.

An alternative and conceptually appealing approach for modeling bounded data is based
on log-symmetric distributions \citep{vanegasp:16a}. This class generalizes the log-normal
distribution by allowing the underlying symmetric kernel to belong to families such as
the normal or Student-$t$, thereby accommodating different tail behaviors and
degrees of robustness. Extending this idea, \citet{Vila2023} introduced the bivariate
log-symmetric (BLS) distribution to model possibly correlated positive random variables.
Building on this construction, \citet{Vila2024} proposed the unit-log-symmetric (ULS)
distribution, obtained through the transformation
\[
W = \frac{T_1}{T_1 + T_2},
\]
where $(T_1,T_2)$ follows a bivariate log-symmetric distribution. This formulation provides
a natural probabilistic foundation for proportion data arising from ratios of positive
quantities.

In many applied settings, such as the proportion of stored hydroelectric energy, the observed bounded
series is naturally expressed as a ratio of latent positive quantities evolving over
time. Ignoring this structural feature may limit interpretability and coherence between
the assumed distribution and the physical mechanism generating the data. These
considerations highlight the need for dynamic quantile models that simultaneously
(i) accommodate asymmetry through conditional quantiles, (ii) capture serial dependence,
and (iii) are grounded on a ratio-based probabilistic construction.

However, to the best of our knowledge, no work has simultaneously combined a ratio-based bounded distribution with ARMA-type dynamics on conditional quantiles. Motivated by these developments, the main objective of this paper is to propose the quantile unit-log-symmetric autoregressive moving average (QULS--ARMA) model, a new class of dynamic models for bounded time series. Secondary objectives are to (a) introduce a quantile reparameterization of the ULS distribution, yielding the QULS distribution; (b) develop conditional maximum likelihood estimation and establish asymptotic properties of the resulting estimators; and (c) illustrate the finite-sample performance of the proposed estimators through Monte Carlo simulations and an empirical application to hydroelectric energy storage proportions.

The rest of this paper proceeds as follows. In Section~\ref{Sec:motivation}, we briefly provide a motivation example. In Section~\ref{Sec:reparam_uls}, we describe the usual ULS distribution and propose
a reparameterization of this distribution in terms of a quantile parameter. In Section~\ref{sec:UQLS-ARMA}, we introduce the QULS--ARMA model, presenting
its probabilistic construction, dynamic specification, and main properties. In this section, we also describe the parameter estimation procedure based on conditional maximum likelihood and discuss inferential aspects. In Section~\ref{Sec:montecarlo}, we report a Monte Carlo simulation
study assessing the finite-sample performance of the proposed estimators under different
scenarios. In Section~\ref{Sec:application}, an empirical application to monthly proportions
of stored hydroelectric energy in Brazil is presented, where the
proposed model is compared with existing alternatives in terms of in-sample fit and
out-of-sample forecasting. Finally, in Section~\ref{Sec:concluding}, we provide some concluding remarks.

\section{Motivating example and modeling rationale}\label{Sec:motivation}

Bounded time series frequently arise in applications where the variable of interest
is naturally expressed as a proportion or ratio. A particularly relevant example in
energy economics and hydrology is the monthly proportion of stored hydroelectric energy \citep{Bayer2017},
which can be interpreted as the ratio between the current volume of stored water and
the total storage capacity of the system. Formally, if $T_{1,t}$ denotes the effective
stored water volume at time $t$ and $T_{2,t}$ the remaining or complementary capacity,
the observed proportion may be written as
\[
Y_t = \frac{T_{1,t}}{T_{1,t}+T_{2,t}}, \qquad 0<Y_t<1.
\]
This representation naturally motivates probability models derived from ratios of
positive random variables, rather than ad hoc bounded distributions.

From a physical and operational perspective, the components $T_{1,t}$ and $T_{2,t}$
are intrinsically linked through the following mechanism: increases in stored volume
necessarily reduce the available remaining capacity, inducing a negative association
between these quantities.  Existing approaches for bounded time series, such as beta, Kumaraswamy and unit Burr XII autoregressive
models, incorporate temporal dependence through the conditional mean or conditional
quantiles, but they do not explicitly reflect the ratio-based structure suggested by
the underlying data-generating mechanism. In contrast, the ULS
distribution arises directly from ratios of log-symmetric random variables, providing a more coherent and interpretable probabilistic foundation for proportion data. It is on this foundation that the proposed QULS--ARMA model is built, extending the ULS distribution to a dynamic framework by allowing autoregressive and moving-average dependence in the conditional quantile.

\section{Reparameterized unit-log-symmetric distribution by the quantile}\label{Sec:reparam_uls}

A continuous random variable $Y$ with support $(0,1)$ is said to follow a unit-log-symmetric (ULS) distribution with parameters $\eta > 0$ and $\sigma > 0$, denoted by $Y\sim {\rm ULS}(\eta, \sigma)$, if its probability density function (PDF) is given by
\begin{equation}\label{eq:uls_pdf_original}
f_Y(y; \eta, \sigma)
=
\frac{1}{\sigma y(1-y)}\,
f_Z\!\left(
\frac{1}{\sigma}\log\!\left[\frac{y}{\eta(1-y)}\right]
\right),
\qquad 0 < y < 1,
\end{equation}
where $f_Z(\cdot)$ denotes the PDF of a standard symmetric random variable $Z$ on the real line, typically following a normal or Student-$t$ distribution.

The corresponding cumulative distribution function (CDF) is
\begin{equation}\label{eq:uls_cdf_original}
F_Y(y; \eta, \sigma)
=
F_Z\!\left(
\frac{1}{\sigma}\log\!\left[\frac{y}{\eta(1-y)}\right]
\right),
\qquad 0 < y < 1,
\end{equation}
where $F_Z(\cdot)$ is the CDF of $Z$.

This transformation maps $(0,1)$ onto $\mathbb{R}$ through
\[
z = \frac{1}{\sigma}\log\!\left(\frac{y}{\eta(1-y)}\right),
\]
ensuring that $f_Y(y; \eta, \sigma)$ is a valid PDF for all $y \in (0,1)$. The parameters $\eta$ and $\sigma$ govern, respectively, the location (through a logit-type shift) and the dispersion of the distribution on the unit interval.


Let $Q_Z(\tau)$ be the $\tau$-th quantile of the base distribution $Z$. From \eqref{eq:uls_cdf_original}, the quantile function of $Y$ is given by
\begin{equation}\label{eq:uls_qtau}
q_{\tau} =
\frac{\eta \exp(\sigma Q_Z(\tau))}{1 + \eta \exp(\sigma Q_Z(\tau))},
\qquad 0 < \tau < 1.
\end{equation}
This ensures $q_{\tau} \in (0,1)$ for all admissible parameters and provides an explicit link between the quantiles of $Z$ and $Y$.

Solving \eqref{eq:uls_qtau} for $\eta$, we obtain
\begin{equation}\label{eq:eta_qtau}
\eta =
\frac{q_{\tau}}{1 - q_{\tau}}
\exp[-\sigma Q_Z(\tau)],
\end{equation}
yielding a one-to-one mapping between $(\eta, \sigma)$ and $(q_{\tau}, \sigma)$ for a fixed quantile level $\tau$.
Hence, we can reparameterize the ULS model in terms of the quantile parameter $q_{\tau}$ instead of $\eta$. By substituting \eqref{eq:eta_qtau} into \eqref{eq:uls_pdf_original}, the PDF of $Y$ under the quantile reparameterization becomes
\begin{equation}\label{eq:uls_pdf_quantile}
f_Y(y; q_{\tau}, \sigma)
=
\frac{1}{\sigma y(1-y)}\,
f_Z\!\left(
\frac{1}{\sigma}
\log\!\left[
\frac{y(1-q_{\tau})\exp(\sigma Q_Z(\tau))}
{q_{\tau}(1-y)}
\right]
\right),
\qquad 0 < y < 1.
\end{equation}
The notation $Y\sim {\rm ULS}(q_{\tau}, \sigma)$ is used.

\section{Quantile unit-log-symmetric ARMA models}\label{sec:UQLS-ARMA}

\subsection{The model}

Let $\{Y_t\}_{t=1}^n$ be a sequence of random variables defined on the probability space $(\Omega,\mathcal{A},\mathbb{P})$ and, for each $t\ge 1$, let
\[
\mathcal{A}_t=\sigma(Y_1,\ldots,Y_t)
=
\sigma\!\big(\{\cap_{i=1}^k Y_i^{-1}(B_i):\, B_1,\ldots,B_k \text{ Borel sets in }\mathbb{R}, \, k=1,\ldots,t\}\big)
\]
be the $\sigma$-algebra generated by the information available up to time $t$, with $\mathcal{A}_0=\{\Omega,\emptyset\}$. Fix a quantile level $\tau\in(0,1)$ and let $Q_Z(\tau)$ denote the $\tau$-th quantile of a symmetric kernel $Z$ on $\mathbb{R}$ (e.g., Normal or Student-$t$). Assume that the conditional distribution of $Y_t$ given $\mathcal{A}_{t-1}$ is unit-log-symmetric, parameterized by its $\tau$-th conditional quantile $q_{\tau,t}\in(0,1)$ and a scale $\sigma>0$:
\[
Y_t \mid \mathcal{A}_{t-1} \sim \mathrm{ULS}\big(q_{\tau,t},\sigma\big).
\]
Then, the conditional PDF of $Y_t \mid \mathcal{A}_{t-1}$ is
\begin{equation}\label{eq:cond_pdf_uls}
f_{Y_t\mid\mathcal{A}_{t-1}}(y_t;\,q_{\tau,t},\sigma)
=
\frac{1}{\sigma\,y_t(1-y_t)}\;
f_Z\!\left(
\frac{1}{\sigma}\log\!\left[\frac{y_t(1-q_{\tau,t})\exp\!\big\{\sigma\,Q_Z(\tau)\big\}}{{q_{\tau,t}}{}(1-y_t)}\right]
\right),
\qquad 0<y_t<1,
\end{equation}
where $f_Z$ is the density of the symmetric kernel $Z$ and $Q_Z(\tau)$ is treated as known for the chosen kernel and $\tau$.

Let $g:(0,1)\to\mathbb{R}$ be a strictly increasing, twice continuously differentiable link (e.g., logit, probit, complementary log–log), with inverse $g^{-1}$. Define the linked conditional quantile
\[
\eta_t \;=\; g(q_{\tau,t}),
\]
and the innovation on the link scale
\[
r_t \;\coloneqq\; g(Y_t)-g(q_{\tau,t}).
\]
Because $g$ is monotone, the conditional $\tau$-quantile commutes with $g$, hence
\[
\mathbb{P}\!\big(r_t\le 0 \mid \mathcal{A}_{t-1}\big)
=\mathbb{P}\!\big(g(Y_t)\le g(q_{\tau,t})\mid \mathcal{A}_{t-1}\big)
=\tau,
\quad\text{so}\quad
Q_{r_t\mid\mathcal{A}_{t-1}}(\tau)=0 \ \ \text{a.s.}
\]

Let $\bm{x}_t=(1,x_{t1},\ldots,x_{tk})^\top$ collect exogenous covariates ($k<n$). We specify
\begin{equation}\label{eq:eta_UQLS}
\eta_t \;=\; g(q_{\tau,t})
\;=\; \alpha + \bm{x}_t^\top\bm{\beta} \;+\; \varrho_t,
\qquad t=m+1,\ldots,n,
\end{equation}
with $m=\max\{p,q\}$ and $\bm{\beta}=(\beta_1,\ldots,\beta_k)^\top$, whereas the dynamic component $\varrho_t$ follows the ARMA scheme
\begin{equation}\label{eq:arma_part_UQLS}
\varrho_t
=\alpha+
\sum_{i=1}^{p}\phi_i\Big[g(Y_{t-i})-\bm{x}_{t-i}^\top\bm{\beta}\Big]
+
\sum_{j=1}^{q}\theta_j\, r_{t-j},
\qquad
Q_{r_t\mid\mathcal{A}_{t-1}}(\tau)=0 \ \ \text{a.s.},
\end{equation}
where $\bm{\phi}=(\phi_1,\ldots,\phi_p)^\top$ and $\bm{\theta}=(\theta_1,\ldots,\theta_q)^\top$ are the autoregressive and moving-average parameter vectors, respectively. Combining \eqref{eq:eta_UQLS}–\eqref{eq:arma_part_UQLS} yields
\begin{equation}\label{eq:UQLS_quantile_rec}
g(q_{\tau,t})
=\alpha+
\bm{x}_t^\top\bm{\beta}
+\sum_{i=1}^{p}\phi_i\Big[g(Y_{t-i})-\bm{x}_{t-i}^\top\bm{\beta}\Big]
+\sum_{j=1}^{q}\theta_j\, r_{t-j},
\qquad
Q_{r_t\mid\mathcal{A}_{t-1}}(\tau)=0.
\end{equation}

Since $g(Y_t)=g(q_{\tau,t})+r_t$, equations \eqref{eq:eta_UQLS} and \eqref{eq:UQLS_quantile_rec} imply
\begin{equation}\label{eq:UQLS_obseq}
g(Y_t)
=\alpha+
\bm{x}_t^\top\bm{\beta}
+\sum_{i=1}^{p}\phi_i\Big[g(Y_{t-i})-\bm{x}_{t-i}^\top\bm{\beta}\Big]
+\sum_{j=1}^{q}\theta_j\, r_{t-j}
+r_t,
\qquad
Q_{r_t\mid\mathcal{A}_{t-1}}(\tau)=0.
\end{equation}
Equation \eqref{eq:UQLS_obseq} defines the QULS--ARMA$(p,q)$ model on the link scale.

Let $L$ denote the lag operator ($L^k X_t=X_{t-k}$) and define the polynomials
\[
\Phi(L)=1-\sum_{i=1}^{p}\phi_i L^i,
\qquad
\Theta(L)=1+\sum_{j=1}^{q}\theta_j L^j.
\]
Then \eqref{eq:UQLS_obseq} can be written compactly as
\begin{equation}\label{eq:lag_form_UQLS}
\Phi(L)\,\Big[g(Y_t)-\bm{x}_t^\top\bm{\beta}\Big]
=
\Theta(L)\, r_t,
\qquad
Q_{r_t\mid\mathcal{A}_{t-1}}(\tau)=0.
\end{equation}

\subsection{Estimation and inference}

Estimation of the parameters of the ULS--ARMA$(p,q)$ model is carried out through
conditional maximum likelihood (CML), using the first $m=\max\{p,q\}$ observations to
initialize the recursion. Let
\[
\boldsymbol{\vartheta}
{
	=
(\vartheta: \vartheta\in\{\alpha,\beta_1,\ldots,\beta_k,\sigma,\phi_1,\ldots,\phi_p,\theta_1,\ldots,\theta_q\})^\top
}
=
(\alpha,\boldsymbol{\beta}^\top, \sigma, \boldsymbol{\phi}^\top, \boldsymbol{\theta}^\top)^\top
\]
denote the full parameter vector, where $\boldsymbol{\beta}$ is the regression coefficient vector,
$\sigma>0$ the scale parameter of the ULS distribution, and $\boldsymbol{\phi}$ and
$\boldsymbol{\theta}$ collect the autoregressive and moving-average parameters, respectively.

Given the conditional density
$f_{Y_t \mid \mathcal{A}_{t-1}}(y_t; q_{\tau,t}, \sigma)$ in \eqref{eq:cond_pdf_uls}, the conditional likelihood based on the sample
$\{y_{m+1},\ldots,y_n\}$ is
\[
L(\boldsymbol{\vartheta})_{m,n}
=
\prod_{t=m+1}^n
f_{Y_t \mid \mathcal{A}_{t-1}}\!\left(
y_t;\, q_{\tau,t}(\boldsymbol{\vartheta}),\, \sigma
\right),
\qquad 0<y_t<1,
\]
which leads to the conditional log-likelihood (up to an additive constant)
\begin{equation}\label{eq:loglik_ULS}
\ell(\boldsymbol{\vartheta})_{m,n}
=
- \sum_{t=m+1}^n \log\!\bigl(\sigma\, y_t(1-y_t)\bigr)
+
\sum_{t=m+1}^n
\log f_Z\!\left(
\frac{1}{\sigma}
\log\!\left[
\frac{y_t(1-q_{\tau,t})\exp\{\sigma Q_Z(\tau)\}}{q_{\tau,t}(1-y_t)}
\right]
\right),
\end{equation}
where $f_Z$ and $Q_Z(\tau)$ are, respectively, the density and $\tau$--quantile of the chosen
symmetric kernel $Z$ (e.g., normal, Student-$t$). Maximization of \eqref{eq:loglik_ULS} is performed numerically. The score vector
$\dot{\ell}(\boldsymbol{\vartheta})$ is obtained by differentiation of
$\ell(\boldsymbol{\vartheta})_{m,n}$ with respect to
each component of $\boldsymbol{\vartheta}$, and the resulting likelihood equations are solved
through iterative methods such as the Broyden--Fletcher--Goldfarb--Shanno (BFGS) algorithm.
{
The components of the score vector
$\dot{\ell}(\boldsymbol{\vartheta})$ are given by
\begin{align*}
	{\partial \ell(\boldsymbol{\vartheta})_{m,n}\over \partial\vartheta}
	=
	- 
	{(n-m)\over \sigma} \,
	\delta_{\vartheta, \sigma}
	+
	\sum_{t=m+1}^n
	{f_Z'(w_t)\over f_Z(w_t)} \,
	{\partial w_t\over \partial \vartheta},
	\quad 
	\vartheta\in\{\alpha,\beta_1,\ldots,\beta_k,\sigma,\phi_1,\ldots,\phi_p,\theta_1,\ldots,\theta_q\},
\end{align*}
where $\delta_{x,y}$ is the dirac delta function and 
\begin{align*}
	w_t
	\equiv 
	\frac{1}{\sigma}
	\log\left[
	\frac{y_t(1-q_{\tau,t})\exp\{\sigma Q_Z(\tau)\}}{q_{\tau,t}(1-y_t)}
	\right].
\end{align*}
Furthermore,
\begin{align} \label{der-1}
&{\partial w_t\over\partial\vartheta}
=
-
\frac{1}{\sigma} \,
\frac{1}{(1-q_{\tau,t})q_{\tau,t}} \,
{\partial q_{\tau,t}\over\partial\vartheta},
\quad 
\vartheta\in\{\alpha,\beta_1,\ldots,\beta_k,\phi_1,\ldots,\phi_p,\theta_1,\ldots,\theta_q\},
\\[0,2cm]
&{\partial w_t\over\partial\sigma}
=
-
\frac{1}{\sigma^2}
\log\left[
\frac{y_t(1-q_{\tau,t})}{q_{\tau,t}(1-y_t)}
\right], \label{der-2}
\end{align}	
and, by using \eqref{eq:UQLS_quantile_rec},
\begin{align}\label{der-q-1}
	&{\partial q_{\tau,t}\over\partial\alpha}
	=
	{1\over g'(q_{\tau,t})}
	\left[
	1
	-
	\sum_{i=1}^{p}\phi_i
	\right],
	\\[0,2cm]
	&{\partial q_{\tau,t}\over\partial\beta_l}
	=
	{1\over g'(q_{\tau,t})} 
	\left[
x_{tl}
	-
	\sum_{i=1}^{p}\phi_i {x}_{(t-i)l}
	\right],  \quad l=1,\ldots,k,  \label{der-q-2}
\\[0,2cm]
	&{\partial q_{\tau,t}\over\partial\phi_u}
=
	{1\over g'(q_{\tau,t})} \,
\Big[g(Y_{t-u})-\bm{x}_{t-u}^\top\bm{\beta}\Big],
\quad u=1,\ldots,p, \label{der-q-3}
\\[0,2cm]
&{\partial q_{\tau,t}\over\partial\theta_v}
=
	{1\over g'(q_{\tau,t})}
\,
\theta_v r_{t-v},
\quad v=1,\ldots,q. \label{der-q-4}
\end{align}	
\noindent
} 
Appropriate starting values are required to initialize the optimization routine; these may be
derived from least-squares-type estimators for the ARMA structure or from standard
\texttt{R} functions such as \texttt{arima}, combined with preliminary estimates of $\sigma$ from ULS models.

In \eqref{eq:loglik_ULS}, when the symmetric kernel is taken to be Student-$t$, the model includes an
additional degrees-of-freedom parameter, denoted by $\nu$. Following \citet{lucas1997}, we do not estimate $\nu$ jointly with the remaining parameters.
As argued in \citet{lucas1997}, the desirable robustness properties of the Student-$t$ specification are
preserved only when $\nu$ is treated as fixed rather than estimated by maximum likelihood. Therefore, the estimation of $\boldsymbol{\vartheta}$ proceeds in two stages for the Student-$t$ case.
\emph{Step 1:} Select a grid of candidate values for $\nu$,
$\nu_1,\nu_2,\ldots,\nu_K$, and for each fixed $\nu_i$ maximize the conditional likelihood with
respect to the remaining parameters, obtaining the corresponding log-likelihood value.
\emph{Step 2:} Choose the value of $\nu_i$ that yields the highest log-likelihood, and adopt the
associated parameter estimates as the final CML estimates.

Under the usual regularity conditions for conditional likelihood estimation
(Assumptions 2.1--2.5 in \citealt{Andersen:70}) and assuming $n$ sufficiently large,
the CML estimator $\widehat{\boldsymbol{\vartheta}}$ is asymptotically normal:
\[
\sqrt{n}\left(\widehat{\boldsymbol{\vartheta}}-\boldsymbol{\vartheta}\right)
\stackrel{\mathcal{D}}{\longrightarrow}{}\;\;
\mathrm{N}_{d}\!\left(
\boldsymbol{0},\,
\mathcal{I}(\boldsymbol{\vartheta})^{-1}
\right),
\qquad n\to\infty,
\]
where $d$ is the dimension of $\boldsymbol{\vartheta}$ and
$\mathcal{I}(\boldsymbol{\vartheta})$ is the Fisher information matrix.
In applications, $\mathcal{I}(\boldsymbol{\vartheta})$ is commonly approximated by the inverse
of the observed information matrix, obtained from the negative Hessian
$\mathcal{J}(\boldsymbol{\vartheta}) = -\partial^2 \ell(\boldsymbol{\vartheta})_{m,n}/
\partial\boldsymbol{\vartheta}\,\partial\boldsymbol{\vartheta}^\top$
evaluated at $\widehat{\boldsymbol{\vartheta}}$.
{ The elements of the Hessian matrix $\partial^2 \ell(\boldsymbol{\vartheta})_{m,n}/\partial\boldsymbol{\vartheta}\,\partial\boldsymbol{\vartheta}^\top$ are given by
\begin{align*}
	{\partial^2 \ell(\boldsymbol{\vartheta})_{m,n}\over \partial\vartheta\partial\widetilde{\vartheta}}
	=
	{(n-m)\over \sigma^2} \,
	\delta_{\vartheta, \sigma}
	\delta_{\widetilde{\vartheta}, \sigma}
	+
	\sum_{t=m+1}^n
	\left[
	{f_Z''(w_t)f_Z(w_t)-\{f_Z'(w_t)\}^2\over \{f_Z(w_t)\}^2} \,
	{\partial w_t\over \partial \vartheta} \,
	{\partial w_t\over \partial \widetilde{\vartheta}}
	+
	{f_Z'(w_t)\over f_Z(w_t)} \,
{\partial^2 w_t\over \partial \vartheta \partial\widetilde{\vartheta}}
	\right],
\end{align*}
where
\begin{align*}
	\vartheta, \widetilde{\vartheta} \in\{\alpha,\beta_1,\ldots,\beta_k,\sigma,\phi_1,\ldots,\phi_p,\theta_1,\ldots,\theta_q\},
\end{align*}
and ${\partial w_t/ \partial \vartheta}$ and
${\partial w_t/ \partial \widetilde{\vartheta}}$ are given in \eqref{der-1}-\eqref{der-2}. Moreover,
\begin{align*}
	&{\partial w_t\over\partial\vartheta}
	=
	\frac{1}{\sigma} \,
		\frac{1}{(1-q_{\tau,t})q_{\tau,t}} 
	\left[
	\frac{(1-2q_{\tau,t}) }{(1-q_{\tau,t})q_{\tau,t}} \,
	{\partial q_{\tau,t}\over\partial\vartheta} \,
	{\partial q_{\tau,t}\over\partial\widetilde{\vartheta}}
	-
	{\partial^2 q_{\tau,t}\over\partial\vartheta\partial\widetilde{\vartheta}}
	\right],
	\ 
	\vartheta,\widetilde{\vartheta}\in\{\alpha,\beta_1,\ldots,\beta_k,\phi_1,\ldots,\phi_p,\theta_1,\ldots,\theta_q\},
	\\[0,2cm]
	&{\partial w_t\over\partial\sigma}
	=
	\frac{2}{\sigma^3}
	\log\left[
	\frac{y_t(1-q_{\tau,t})}{q_{\tau,t}(1-y_t)}
	\right],
\end{align*}	
where 	${\partial q_{\tau,t}/\partial\vartheta}$ and
${\partial q_{\tau,t}/\partial\widetilde{\vartheta}}$ are given in \eqref{der-q-1}-\eqref{der-q-4},
and 	${\partial^2 q_{\tau,t}/\partial\vartheta\partial\widetilde{\vartheta}}$ can be readily derived through routine calculations from \eqref{der-q-1}-\eqref{der-q-4}.
}

The quantile level $\tau$ is chosen by the practitioner according to the
purpose of the analysis, and the selected $\tau$ determines the conditional quantile tracked by the
ULS--ARMA model. Regarding the link $g(\cdot)$, common options include the logit, probit, and
complementary log--log functions. Model selection can be guided by information criteria such as
AIC or BIC. The logit link is particularly attractive because it ensures that
$q_{\tau,t}\in(0,1)$ for all $t$.

As in other quantile-based time series frameworks, if several quantiles are modeled independently,
their estimated trajectories might not satisfy the natural monotonicity across $\tau$.
Hence, care must be exercised when interpreting multiple-quantile ULS--ARMA fits, as
pointed out by \citet{koenkerxiao:2006} in the broader context of quantile regression.

\subsection{Prediction}

After obtaining the conditional maximum likelihood estimates of the parameters in the
ULS--ARMA$(p,q)$ model, we now describe the procedure for generating forecasts of the bounded
time series $\{Y_t\}$. Let $\widehat{Y}_{t+h}$ denote the $h$-step-ahead prediction based on the
information set $\mathcal{A}_t$. For notational convenience, we set
\[
\widehat{Y}_{t+h} =
\begin{cases}
\widehat{Y}_t(h), & h > 0,\\[0.15cm]
Y_{t+h}, & h \le 0,
\end{cases}
\qquad\text{and}\qquad
\widehat{r}_{t+h} =
\begin{cases}
0, & h > 0,\\[0.15cm]
\widehat{r}_{t+h}, & h \le 0,
\end{cases}
\]
reflecting the fact that future innovations on the link scale have conditional $\tau$--quantile equal to zero. The conditional quantile is determined through the link function $g:(0,1)\to\mathbb{R}$ and the ARMA
structure imposed on $g(q_{\tau,t})$. Given the estimated parameters
$\widehat{\boldsymbol{\beta}}$, $\widehat{\phi}_1,\ldots,\widehat{\phi}_p$ and
$\widehat{\theta}_1,\ldots,\widehat{\theta}_q$, the fitted conditional $\tau$--quantile at time $t$ is
\begin{equation}\label{eq:predQhat_ULS}
\widehat{\eta}_t
= g(\widehat{q}_{\tau,t})
=
\widehat\alpha+\boldsymbol{x}_t^\top \widehat{\boldsymbol{\beta}}
+
\sum_{i=1}^p \widehat{\phi}_i \left[g(Y_{t-i}) - \boldsymbol{x}_{t-i}^\top \widehat{\boldsymbol{\beta}} \right]
+
\sum_{j=1}^q \widehat{\theta}_j\, \widehat{r}_{t-j}.
\end{equation}

Since the innovation on the link scale satisfies
\[
r_t = g(Y_t) - g(q_{\tau,t}),
\]
an empirical estimate is obtained from the fitted model as
\[
\widehat{r}_t = g(Y_t) - \widehat{\eta}_t,
\qquad t = m+1,\ldots,n,
\]
which justifies the convention $\widehat{r}_{t+h}=0$ for $h>0$.

Using \eqref{eq:predQhat_ULS}, the one-step-ahead forecast is given by
\begin{equation}
\widehat{Y}_{n+1}
=
g^{-1}\!\left(
\widehat\alpha+\boldsymbol{x}_{n+1}^\top \widehat{\boldsymbol{\beta}}
+
\sum_{i=1}^{p} \widehat{\phi}_i \left[g(Y_{n+1-i}) - \boldsymbol{x}_{n+1-i}^\top \widehat{\boldsymbol{\beta}}\right]
+
\sum_{j=1}^{q} \widehat{\theta}_j \widehat{r}_{n+1-j}
\right).
\end{equation}

The same reasoning yields the forecast for time $n+2$:
\[
\widehat{Y}_{n+2}
=
g^{-1}\!\left(
\widehat\alpha+\boldsymbol{x}_{n+2}^\top \widehat{\boldsymbol{\beta}}
+
\sum_{i=1}^{p} \widehat{\phi}_i \left[g(\widehat{Y}_{n+2-i}) - \boldsymbol{x}_{n+2-i}^\top \widehat{\boldsymbol{\beta}}\right]
+
\sum_{j=1}^{q} \widehat{\theta}_j \widehat{r}_{n+2-j}
\right),
\]
where all terms involving future values are replaced by their corresponding forecasts. The recursion
extends naturally to any horizon $h>2$.

\section{Monte Carlo simulation study}
\label{Sec:montecarlo}

This section reports a Monte Carlo study assessing the finite-sample performance of the
conditional maximum likelihood (CML) estimators in the proposed ULS--ARMA framework under the normal kernel. The objective is to assess bias, relative bias, absolute
relative bias, and root mean squared error of the estimators across different dependence structures,
dispersion levels, and quantile levels $\tau\in\{0.25,0.50,0.75\}$.

\subsection{Data generating process}

Let $\{Y_t\}_{t\ge 1}$ be a bounded time series with $Y_t\in(0,1)$. Fix a quantile level
$\tau\in(0,1)$ and define the linked conditional quantile
$\eta_t=g(q_{\tau,t})$ with $g(\cdot)$ the logit link,
$g(u)=\log\{u/(1-u)\}$.
Under the ULS--ARMA$(p,q)$ model,
\begin{equation}\label{eq:mc_eta}
\eta_t
=
\alpha + \bm{x}_t^\top\bm{\beta}
+
\sum_{i=1}^{p}\phi_i\Big(g(Y_{t-i})-\bm{x}_{t-i}^\top\bm{\beta}\Big)
+
\sum_{j=1}^{q}\theta_j r_{t-j},
\qquad t>m,
\end{equation}
where $m=\max\{p,q\}$, $\bm{\beta}=(\beta_1,\beta_2)^\top$, and the innovation on the link scale is
$r_t=g(Y_t)-\eta_t$.
The conditional distribution is assumed to be unit-log-symmetric,
\begin{equation}\label{eq:mc_cond}
Y_t \mid \mathcal{A}_{t-1} \sim \mathrm{ULS}(q_{\tau,t},\sigma;\text{kernel}), \qquad
q_{\tau,t}=g^{-1}(\eta_t), \qquad \sigma>0,
\end{equation}
where ``kernel'' stands for Normal$(0,1)$.

In order to simulate from \eqref{eq:mc_cond} under the quantile parameterization, we exploit the explicit stochastic representation induced by the logit transformation.
Let $Z_t$ be drawn independently from the chosen kernel with $\tau$-th quantile $Q_Z(\tau)$.
Then
\begin{equation}\label{eq:mc_representation}
Y_t
=
g^{-1}\!\Big(\eta_t+\sigma\{Z_t-Q_Z(\tau)\}\Big),
\end{equation}
which ensures that the conditional $\tau$-quantile of $Y_t$ given $\mathcal{A}_{t-1}$ is exactly
$q_{\tau,t}$, since $\mathbb{P}(Z_t\le Q_Z(\tau))=\tau$ by construction.
For the normal kernel, $Q_Z(\tau)=\Phi^{-1}(\tau)$.
The simulated innovation $r_t$ follows immediately as $r_t=g(Y_t)-\eta_t$.
We consider three quantile levels, $\tau\in\{0.25, 0.50, 0.75\}$, in order to assess the
finite-sample performance of the CML estimators across the lower quartile, the conditional
median, and the upper quartile of the conditional distribution.

We adopt a parsimonious harmonic specification to generate deterministic covariates,
\[
\bm{x}_t=\big(\cos(2\pi t/12),\,\sin(2\pi t/12)\big)^\top,
\]
so that $\bm{x}_t^\top\bm{\beta}$ acts as a seasonal component on the link scale. This is aligned
with the empirical application, where harmonic terms capture annual periodicity. We consider four scenarios combining different dynamic orders and dispersion
levels. The parameter vectors are set to:
\[
\text{S1: } (p,q)=(2,0),\;
(\alpha,\bm{\beta},\sigma,\bm{\phi})
=
(0.50,(0.50,0.20),0.10,(1.20,-0.30)),
\]
\[
\text{S2: } (p,q)=(2,0),\;
(\alpha,\bm{\beta},\sigma,\bm{\phi})
=
(0.10,(0.50,0.20),0.20,(1.20,-0.30)),
\]
\[
\text{S3: } (p,q)=(1,1),\;
(\alpha,\bm{\beta},\sigma,\phi_1,\theta_1)
=
(0.40,(0.50,0.20),0.10,0.85,0.20),
\]
\[
\text{S4: } (p,q)=(1,1),\;
(\alpha,\bm{\beta},\sigma,\phi_1,\theta_1)
=
(0.90,(0.50,0.20),0.20,0.85,0.20).
\]
Scenarios S1--S2 isolate the effect of dispersion ($\sigma$) under an autoregressive structure,
whereas S3--S4 assess the impact of a moving-average component under low/high dispersion. For each scenario, we consider sample sizes $n\in\{75,125,200,400\}$ and perform $R$ Monte Carlo
replications. In each replication, a burn-in period of length $B$ is used to mitigate
initialization effects, and only the last $n$ observations are retained for estimation.

\subsection{Performance measures}

For each replication, the parameter vector is estimated by maximizing the conditional
log-likelihood $\ell(\boldsymbol{\vartheta})_{m,n}$ defined in
\eqref{eq:loglik_ULS}, with $\boldsymbol{\vartheta}=(\alpha,\bm{\beta}^\top,\sigma,\bm{\phi}^\top,\bm{\theta}^\top)^\top$. The recursion for $q_{\tau,t}$ is computed from \eqref{eq:mc_eta} using the observed past values
$\{Y_{t-i}\}$ and the fitted residuals $\widehat{r}_{t-j}$, initialized at zero for the first
$m$ observations. Numerical optimization is carried out via BFGS.

Let $\widehat{\psi}^{(r)}$ denote the estimate of a generic parameter $\psi$ at replication
$r=1,\ldots,R$, and let $\psi_0$ be its true value. We compute:
\begin{align}
\mathrm{RB}(\widehat{\psi})
&=
 \frac{\frac{1}{R}\sum_{r=1}^R(\widehat{\psi}^{(r)}-\psi_0)}{\psi_0},
\label{eq:mc_rb}\\[0.15cm]
\mathrm{ARB}(\widehat{\psi})
&=
 \frac{\frac{1}{R}\sum_{r=1}^R\left|\widehat{\psi}^{(r)}-\psi_0\right|}{|\psi_0|},
\label{eq:mc_arb}\\[0.15cm]
\mathrm{RMSE}(\widehat{\psi})
&=
\left\{\frac{1}{R}\sum_{r=1}^R(\widehat{\psi}^{(r)}-\psi_0)^2\right\}^{1/2}.
\label{eq:mc_rmse}
\end{align}
These measures are reported for each parameter and $(n,\text{scenario})$ combination. Algorithm~\ref{alg:mc_uls_normal} summarizes the simulation and estimation steps.

\begin{algorithm}[!ht]
\caption{Monte Carlo experiment for the ULS--ARMA model (normal kernel).}
\label{alg:mc_uls_normal}
\begin{enumerate}
\item Fix $\tau\in\{0.25,0.50,0.75\}$, sample sizes $\mathcal{N}=\{75,125,200,400\}$, replications $R$, burn-in $B$, and the set of scenarios $\{\text{S1},\ldots,\text{S4}\}$.
\item For each scenario S$\in\{\text{S1},\ldots,\text{S4}\}$, each $n\in\mathcal{N}$, and each $\tau$:
\begin{enumerate}
\item For $r=1,\ldots,R$:
\begin{enumerate}
\item Initialize $Y_1,\ldots,Y_m\in(0,1)$, set $\widehat{r}_1=\cdots=\widehat{r}_m=0$, and define $\bm{x}_t=(\cos(2\pi t/12),\sin(2\pi t/12))^\top$.
\item For $t=m+1,\ldots,n+B$:
\begin{enumerate}
\item Compute $\eta_t$ from \eqref{eq:mc_eta} using the scenario parameters.
\item Set $q_{\tau,t}=g^{-1}(\eta_t)$ and draw $Z_t$ from the normal kernel.
\item Generate $Y_t=g^{-1}\!\big(\eta_t+\sigma\{Z_t-Q_Z(\tau)\}\big)$ as in \eqref{eq:mc_representation}.
\item Update $r_t=g(Y_t)-\eta_t$.
\end{enumerate}
\item Discard the first $B$ observations and retain $Y_{B+1},\ldots,Y_{B+n}$.
\item Fit the ULS--ARMA model by CML under the normal kernel. Store $\widehat{\boldsymbol{\vartheta}}^{(r)}=(\widehat{\alpha},\widehat{\bm{\beta}},\widehat{\bm{\phi}},\widehat{\bm{\theta}},\widehat{\sigma})^\top$.
\end{enumerate}
\item Compute $\mathrm{RB}$, $\mathrm{ARB}$, and $\mathrm{RMSE}$ via \eqref{eq:mc_rb}--\eqref{eq:mc_rmse} for all parameters.
\end{enumerate}
\end{enumerate}
\end{algorithm}

\subsection{Results}


The Monte Carlo results for the normal kernel at $\tau=0.50$ are reported in Tables~\ref{tab:mc_rb_normal_tau050}--\ref{tab:mc_rmse_normal_tau050}. From these tables, we note that, across all scenarios, the
estimators exhibit the expected improvement in accuracy as the sample size increases, with
systematic reductions in relative bias (RB), absolute relative bias (ARB) and root mean squared error (RMSE) as
$n$ grows.

In scenarios S1 and S2, which correspond to purely autoregressive dynamics, the regression
coefficients $\beta_1$ and $\beta_2$ display very small relative bias even for the smallest
sample size and become essentially unbiased for $n\ge 200$. The autoregressive parameters
$\phi_1$ and $\phi_2$ present moderate negative relative bias in small samples. However, both ARB and RMSE for
these parameters decrease markedly as $n$ increases, indicating consistency of the estimators.
The dispersion parameter $\sigma$ is mildly downward biased in relative terms, but its ARB and
RMSE remain small throughout, showing stable estimation under both low (S1) and moderate (S2)
dispersion levels.

Scenarios S3 and S4 introduce a moving-average component. In these cases, the MA parameter $\theta_1$ exhibits comparatively
large relative bias and ARB for small sample sizes. Nevertheless, both ARB and RMSE for
$\theta_1$ decrease as the sample size increases, suggesting that the estimator is consistent.
The autoregressive coefficient $\phi_1$ and the regression parameters maintain good finite-sample
properties, with bias and variability comparable to those observed in scenarios S1--S2. The
estimation of the dispersion parameter $\sigma$ is more sensitive in scenario S3, where the true
dispersion is low, but improves substantially in scenario S4 when the true $\sigma$ is larger.

The RMSE results further corroborate these findings by showing a clear decline as $n$ increases
for all parameters. Overall, the Monte Carlo evidence supports the reliability of
the proposed ULS--ARMA estimators across a wide
range of dynamic structures and dispersion levels.


Tables~\ref{tab:mc_rb_tau025}--\ref{tab:mc_rmse_tau025} and Tables~\ref{tab:mc_rb_tau075}--\ref{tab:mc_rmse_tau075} report
the corresponding results for $\tau=0.25$ and $\tau=0.75$, respectively. The qualitative behavior of the estimators is broadly consistent with the findings for $\tau=0.50$: bias and variability decrease as $n$ increases for all parameters, whereas the regression coefficients $\beta_1$ and $\beta_2$ remain well-estimated even in small samples. The scale parameter $\sigma$ tends to display somewhat larger relative bias at $\tau=0.25$ compared with $\tau=0.75$, reflecting greater sampling variability in the lower tail of the conditional distribution. In scenarios S3 and S4, the MA coefficient $\theta_1$ retains its characteristic larger bias in small samples regardless of $\tau$, though the rate of improvement with $n$ is similar across quantile levels. These results confirm that the finite-sample performance of the CML estimators is robust to the choice of $\tau$ within the range $[0.25, 0.75]$.

\begin{table}[!ht]
\centering
\caption{\label{tab:mc_rb_normal_tau050}
Monte Carlo results (normal kernel, $\tau=0.50$): relative bias (RB) with scenarios S1--S4.}
\fontsize{9}{11}\selectfont
\begin{tabular}[t]{llcccccccc}
\toprule
 & & \multicolumn{4}{c}{S1} & \multicolumn{4}{c}{S2} \\
\cmidrule(l{3pt}r{3pt}){3-6} \cmidrule(l{3pt}r{3pt}){7-10}
 & Parameter & $n=75$ & $n=125$ & $n=200$ & $n=400$ & $n=75$ & $n=125$ & $n=200$ & $n=400$\\
\midrule
 & $\alpha$    &  0.4060 &  0.1930 &  0.1144 &  0.0434 &  0.3808 &  0.1985 &  0.1159 &  0.0500 \\
 & $\beta_{1}$ &  0.0066 &  0.0029 & -0.0057 &  0.0002 &  0.0131 &  0.0059 & -0.0113 &  0.0004 \\
 & $\beta_{2}$ &  0.0013 &  0.0178 &  0.0006 & -0.0081 &  0.0027 &  0.0355 &  0.0013 & -0.0163 \\
 & $\phi_{1}$  & -0.0567 & -0.0287 & -0.0142 & -0.0102 & -0.0567 & -0.0287 & -0.0142 & -0.0102 \\
 & $\phi_{2}$  & -0.0904 & -0.0506 & -0.0189 & -0.0266 & -0.0904 & -0.0506 & -0.0189 & -0.0266 \\
 & $\sigma$    & -0.0389 & -0.0200 & -0.0100 & -0.0084 & -0.0389 & -0.0200 & -0.0100 & -0.0084 \\
\midrule
 & & \multicolumn{4}{c}{S3} & \multicolumn{4}{c}{S4} \\
\cmidrule(l{3pt}r{3pt}){3-6} \cmidrule(l{3pt}r{3pt}){7-10}
 & Parameter & $n=75$ & $n=125$ & $n=200$ & $n=400$ & $n=75$ & $n=125$ & $n=200$ & $n=400$\\
\midrule
 & $\alpha$     &  0.3297 &  0.2550 &  0.0951 &  0.0538 &  0.2268 &  0.1878 &  0.0501 &  0.0290 \\
 & $\beta_{1}$  & -0.0030 &  0.0104 & -0.0064 & -0.0034 & -0.0102 &  0.0087 & -0.0190 & -0.0072 \\
 & $\beta_{2}$  &  0.0144 & -0.0260 & -0.0349 &  0.0044 &  0.0092 & -0.0908 & -0.0774 & -0.0074 \\
 & $\phi_{1}$   & -0.0597 & -0.0445 & -0.0173 & -0.0099 & -0.0409 & -0.0327 & -0.0089 & -0.0055 \\
 & $\sigma$     &  0.9424 &  0.9670 &  0.9746 &  0.9811 &  0.0111 &  0.0062 &  0.0038 &  0.0328 \\
 & $\theta_{1}$ & -0.5086 & -0.5027 & -0.5077 & -0.5049 & -0.0515 & -0.0434 & -0.0475 & -0.0046 \\
\bottomrule
\end{tabular}
\end{table}

\begin{table}[!ht]
\centering
\caption{\label{tab:mc_arb_normal_tau050}
Monte Carlo results (normal kernel, $\tau=0.50$): absolute relative bias (ARB) with scenarios S1--S4.}
\fontsize{9}{11}\selectfont
\begin{tabular}[t]{llcccccccc}
\toprule
 & & \multicolumn{4}{c}{S1} & \multicolumn{4}{c}{S2} \\
\cmidrule(l{3pt}r{3pt}){3-6} \cmidrule(l{3pt}r{3pt}){7-10}
 & Parameter & $n=75$ & $n=125$ & $n=200$ & $n=400$ & $n=75$ & $n=125$ & $n=200$ & $n=400$\\
\midrule
 & $\alpha$    & 0.5404 & 0.3151 & 0.2203 & 0.1593 & 0.5469 & 0.3578 & 0.2517 & 0.1862 \\
 & $\beta_{1}$ & 0.0717 & 0.0618 & 0.0427 & 0.0324 & 0.1433 & 0.1235 & 0.0854 & 0.0649 \\
 & $\beta_{2}$ & 0.1960 & 0.1325 & 0.1100 & 0.0792 & 0.3920 & 0.2649 & 0.2199 & 0.1584 \\
 & $\phi_{1}$  & 0.0883 & 0.0647 & 0.0503 & 0.0322 & 0.0883 & 0.0647 & 0.0503 & 0.0322 \\
 & $\phi_{2}$  & 0.3102 & 0.2323 & 0.1931 & 0.1318 & 0.3102 & 0.2323 & 0.1931 & 0.1318 \\
 & $\sigma$    & 0.0697 & 0.0561 & 0.0386 & 0.0306 & 0.0697 & 0.0561 & 0.0386 & 0.0306 \\
\midrule
 & & \multicolumn{4}{c}{S3} & \multicolumn{4}{c}{S4} \\
\cmidrule(l{3pt}r{3pt}){3-6} \cmidrule(l{3pt}r{3pt}){7-10}
 & Parameter & $n=75$ & $n=125$ & $n=200$ & $n=400$ & $n=75$ & $n=125$ & $n=200$ & $n=400$\\
\midrule
 & $\alpha$     & 0.4781 & 0.3728 & 0.2352 & 0.1694 & 0.4433 & 0.3464 & 0.2376 & 0.1744 \\
 & $\beta_{1}$  & 0.1540 & 0.1166 & 0.0900 & 0.0655 & 0.2362 & 0.1869 & 0.1468 & 0.1076 \\
 & $\beta_{2}$  & 0.3524 & 0.2984 & 0.2380 & 0.1654 & 0.4934 & 0.4261 & 0.3595 & 0.2482 \\
 & $\phi_{1}$   & 0.0834 & 0.0642 & 0.0408 & 0.0296 & 0.0776 & 0.0608 & 0.0415 & 0.0309 \\
 & $\sigma$     & 0.9424 & 0.9670 & 0.9746 & 0.9811 & 0.0793 & 0.0618 & 0.0492 & 0.0541 \\
 & $\theta_{1}$ & 0.5086 & 0.5027 & 0.5077 & 0.5049 & 0.0996 & 0.0692 & 0.0599 & 0.0599 \\
\bottomrule
\end{tabular}
\end{table}

\begin{table}[!ht]
\centering
\caption{\label{tab:mc_rmse_normal_tau050}
Monte Carlo results (normal kernel, $\tau=0.50$): RMSE with scenarios S1--S4.}
\fontsize{9}{11}\selectfont
\begin{tabular}[t]{llcccccccc}
\toprule
 & & \multicolumn{4}{c}{S1} & \multicolumn{4}{c}{S2} \\
\cmidrule(l{3pt}r{3pt}){3-6} \cmidrule(l{3pt}r{3pt}){7-10}
 & Parameter & $n=75$ & $n=125$ & $n=200$ & $n=400$ & $n=75$ & $n=125$ & $n=200$ & $n=400$\\
\midrule
 & $\alpha$    & 0.3576 & 0.2131 & 0.1449 & 0.1020 & 0.0731 & 0.0481 & 0.0329 & 0.0237 \\
 & $\beta_{1}$ & 0.0446 & 0.0381 & 0.0267 & 0.0203 & 0.0893 & 0.0762 & 0.0534 & 0.0406 \\
 & $\beta_{2}$ & 0.0484 & 0.0335 & 0.0277 & 0.0196 & 0.0968 & 0.0670 & 0.0553 & 0.0392 \\
 & $\phi_{1}$  & 0.1360 & 0.0992 & 0.0766 & 0.0497 & 0.1360 & 0.0992 & 0.0766 & 0.0497 \\
 & $\phi_{2}$  & 0.1167 & 0.0884 & 0.0730 & 0.0495 & 0.1167 & 0.0884 & 0.0730 & 0.0495 \\
 & $\sigma$    & 0.0089 & 0.0071 & 0.0047 & 0.0038 & 0.0178 & 0.0142 & 0.0095 & 0.0076 \\
\midrule
 & & \multicolumn{4}{c}{S3} & \multicolumn{4}{c}{S4} \\
\cmidrule(l{3pt}r{3pt}){3-6} \cmidrule(l{3pt}r{3pt}){7-10}
 & Parameter & $n=75$ & $n=125$ & $n=200$ & $n=400$ & $n=75$ & $n=125$ & $n=200$ & $n=400$\\
\midrule
 & $\alpha$     & 0.2580 & 0.1998 & 0.1196 & 0.0852 & 0.5473 & 0.4122 & 0.2777 & 0.2080 \\
 & $\beta_{1}$  & 0.0945 & 0.0725 & 0.0576 & 0.0403 & 0.1449 & 0.1197 & 0.0921 & 0.0702 \\
 & $\beta_{2}$  & 0.0895 & 0.0730 & 0.0595 & 0.0418 & 0.1247 & 0.1052 & 0.0927 & 0.0662 \\
 & $\phi_{1}$   & 0.0963 & 0.0730 & 0.0434 & 0.0318 & 0.0913 & 0.0682 & 0.0451 & 0.0344 \\
 & $\sigma$     & 0.0958 & 0.0975 & 0.0979 & 0.0984 & 0.0204 & 0.0156 & 0.0128 & 0.0207 \\
 & $\theta_{1}$ & 0.1045 & 0.1024 & 0.1025 & 0.1016 & 0.0253 & 0.0175 & 0.0148 & 0.0219 \\
\bottomrule
\end{tabular}
\end{table}

\begin{table}[!ht]
\centering
\caption{\label{tab:mc_rb_tau025}
Monte Carlo results (normal kernel, $\tau=0.25$): relative bias (RB) with scenarios S1--S4.}
\fontsize{9}{11}\selectfont
\begin{tabular}[t]{llcccccccc}
\toprule
 & & \multicolumn{4}{c}{S1} & \multicolumn{4}{c}{S2} \\
\cmidrule(l{3pt}r{3pt}){3-6} \cmidrule(l{3pt}r{3pt}){7-10}
 & Parameter & $n=75$ & $n=125$ & $n=200$ & $n=400$ & $n=75$ & $n=125$ & $n=200$ & $n=400$\\
\midrule
 & $\alpha$    &  0.4664 &  0.2217 &  0.1311 &  0.0503 &  0.9847 &  0.4850 &  0.2834 &  0.1189 \\
 & $\beta_{1}$ &  0.0066 &  0.0029 & -0.0057 &  0.0002 &  0.0131 &  0.0059 & -0.0113 &  0.0004 \\
 & $\beta_{2}$ &  0.0013 &  0.0178 &  0.0006 & -0.0081 &  0.0027 &  0.0355 &  0.0012 & -0.0163 \\
 & $\phi_{1}$  & -0.0567 & -0.0287 & -0.0142 & -0.0102 & -0.0567 & -0.0287 & -0.0142 & -0.0102 \\
 & $\phi_{2}$  & -0.0904 & -0.0506 & -0.0189 & -0.0266 & -0.0904 & -0.0506 & -0.0189 & -0.0266 \\
 & $\sigma$    & -0.0389 & -0.0200 & -0.0100 & -0.0084 & -0.0389 & -0.0200 & -0.0100 & -0.0084 \\
\midrule
 & & \multicolumn{4}{c}{S3} & \multicolumn{4}{c}{S4} \\
\cmidrule(l{3pt}r{3pt}){3-6} \cmidrule(l{3pt}r{3pt}){7-10}
 & Parameter & $n=75$ & $n=125$ & $n=200$ & $n=400$ & $n=75$ & $n=125$ & $n=200$ & $n=400$\\
\midrule
 & $\alpha$     &  0.2278 &  0.1344 & -0.0527 & -0.1021 &  0.2581 &  0.2083 &  0.0530 &  0.0319 \\
 & $\beta_{1}$  & -0.0030 &  0.0104 & -0.0064 & -0.0034 &  0.0010 &  0.0285 & -0.0126 & -0.0091 \\
 & $\beta_{2}$  &  0.0144 & -0.0260 & -0.0349 &  0.0044 & -0.0129 & -0.0530 & -0.0575 & -0.0031 \\
 & $\phi_{1}$   & -0.0597 & -0.0445 & -0.0173 & -0.0099 & -0.0405 & -0.0323 & -0.0092 & -0.0059 \\
 & $\sigma$     &  0.9424 &  0.9670 &  0.9746 &  0.9811 &  0.0052 &  0.0270 &  0.0392 &  0.0289 \\
 & $\theta_{1}$ & -0.5086 & -0.5027 & -0.5077 & -0.5049 & -0.0573 & -0.0223 & -0.0109 & -0.0098 \\
\bottomrule
\end{tabular}
\end{table}

\begin{table}[!ht]
\centering
\caption{\label{tab:mc_rmse_tau025}
Monte Carlo results (normal kernel, $\tau=0.25$): RMSE with scenarios S1--S4.}
\fontsize{9}{11}\selectfont
\begin{tabular}[t]{llcccccccc}
\toprule
 & & \multicolumn{4}{c}{S1} & \multicolumn{4}{c}{S2} \\
\cmidrule(l{3pt}r{3pt}){3-6} \cmidrule(l{3pt}r{3pt}){7-10}
 & Parameter & $n=75$ & $n=125$ & $n=200$ & $n=400$ & $n=75$ & $n=125$ & $n=200$ & $n=400$\\
\midrule
 & $\alpha$    & 0.4083 & 0.2425 & 0.1650 & 0.1157 & 0.1707 & 0.1046 & 0.0710 & 0.0501 \\
 & $\beta_{1}$ & 0.0446 & 0.0381 & 0.0267 & 0.0203 & 0.0893 & 0.0762 & 0.0534 & 0.0406 \\
 & $\beta_{2}$ & 0.0484 & 0.0335 & 0.0277 & 0.0196 & 0.0968 & 0.0670 & 0.0553 & 0.0392 \\
 & $\phi_{1}$  & 0.1360 & 0.0992 & 0.0766 & 0.0497 & 0.1360 & 0.0992 & 0.0766 & 0.0497 \\
 & $\phi_{2}$  & 0.1167 & 0.0884 & 0.0730 & 0.0495 & 0.1167 & 0.0884 & 0.0730 & 0.0495 \\
 & $\sigma$    & 0.0089 & 0.0071 & 0.0047 & 0.0038 & 0.0178 & 0.0142 & 0.0095 & 0.0076 \\
\midrule
 & & \multicolumn{4}{c}{S3} & \multicolumn{4}{c}{S4} \\
\cmidrule(l{3pt}r{3pt}){3-6} \cmidrule(l{3pt}r{3pt}){7-10}
 & Parameter & $n=75$ & $n=125$ & $n=200$ & $n=400$ & $n=75$ & $n=125$ & $n=200$ & $n=400$\\
\midrule
 & $\alpha$     & 0.2744 & 0.2080 & 0.1332 & 0.1044 & 0.6297 & 0.4905 & 0.3317 & 0.2290 \\
 & $\beta_{1}$  & 0.0945 & 0.0725 & 0.0576 & 0.0403 & 0.1448 & 0.1135 & 0.0945 & 0.0687 \\
 & $\beta_{2}$  & 0.0895 & 0.0730 & 0.0595 & 0.0418 & 0.1216 & 0.1082 & 0.0879 & 0.0601 \\
 & $\phi_{1}$   & 0.0963 & 0.0730 & 0.0434 & 0.0318 & 0.0914 & 0.0708 & 0.0469 & 0.0336 \\
 & $\sigma$     & 0.0958 & 0.0975 & 0.0979 & 0.0984 & 0.0214 & 0.0220 & 0.0297 & 0.0178 \\
 & $\theta_{1}$ & 0.1045 & 0.1024 & 0.1025 & 0.1016 & 0.0256 & 0.0251 & 0.0321 & 0.0190 \\
\bottomrule
\end{tabular}
\end{table}

\begin{table}[!ht]
\centering
\caption{\label{tab:mc_rb_tau075}
Monte Carlo results (normal kernel, $\tau=0.75$): relative bias (RB) with scenarios S1--S4.}
\fontsize{9}{11}\selectfont
\begin{tabular}[t]{llcccccccc}
\toprule
 & & \multicolumn{4}{c}{S1} & \multicolumn{4}{c}{S2} \\
\cmidrule(l{3pt}r{3pt}){3-6} \cmidrule(l{3pt}r{3pt}){7-10}
 & Parameter & $n=75$ & $n=125$ & $n=200$ & $n=400$ & $n=75$ & $n=125$ & $n=200$ & $n=400$\\
\midrule
 & $\alpha$    & 0.3456 & 0.1644 & 0.0976 & 0.0365 & $-$0.2231 & $-$0.0880 & $-$0.0517 & $-$0.0189 \\
 & $\beta_{1}$ & 0.0066 & 0.0029 & $-$0.0057 & 0.0002 & 0.0131 & 0.0059 & $-$0.0113 & 0.0004 \\
 & $\beta_{2}$ & 0.0013 & 0.0178 & 0.0006 & $-$0.0081 & 0.0027 & 0.0355 & 0.0012 & $-$0.0163 \\
 & $\phi_{1}$  & $-$0.0567 & $-$0.0287 & $-$0.0142 & $-$0.0102 & $-$0.0567 & $-$0.0287 & $-$0.0142 & $-$0.0102 \\
 & $\phi_{2}$  & $-$0.0904 & $-$0.0506 & $-$0.0189 & $-$0.0266 & $-$0.0904 & $-$0.0506 & $-$0.0189 & $-$0.0266 \\
 & $\sigma$    & $-$0.0389 & $-$0.0200 & $-$0.0100 & $-$0.0084 & $-$0.0389 & $-$0.0200 & $-$0.0100 & $-$0.0084 \\
\midrule
 & & \multicolumn{4}{c}{S3} & \multicolumn{4}{c}{S4} \\
\cmidrule(l{3pt}r{3pt}){3-6} \cmidrule(l{3pt}r{3pt}){7-10}
 & Parameter & $n=75$ & $n=125$ & $n=200$ & $n=400$ & $n=75$ & $n=125$ & $n=200$ & $n=400$\\
\midrule
 & $\alpha$     & 0.4317 & 0.3755 & 0.2430 & 0.2098 & 0.1803 & 0.1803 & 0.0657 & 0.0268 \\
 & $\beta_{1}$  & $-$0.0019 & 0.0104 & $-$0.0064 & $-$0.0034 & $-$0.0056 & 0.0038 & $-$0.0143 & $-$0.0070 \\
 & $\beta_{2}$  & 0.0180 & $-$0.0260 & $-$0.0349 & 0.0044 & $-$0.0125 & $-$0.0448 & $-$0.0409 & 0.0059 \\
 & $\phi_{1}$   & $-$0.0597 & $-$0.0445 & $-$0.0173 & $-$0.0099 & $-$0.0382 & $-$0.0360 & $-$0.0133 & $-$0.0053 \\
 & $\sigma$     & 0.9430 & 0.9670 & 0.9746 & 0.9811 & 0.0097 & 0.0217 & 0.0216 & 0.0188 \\
 & $\theta_{1}$ & $-$0.5091 & $-$0.5027 & $-$0.5077 & $-$0.5049 & $-$0.0489 & $-$0.0290 & $-$0.0262 & $-$0.0191 \\
\bottomrule
\end{tabular}
\end{table}

\clearpage

\begin{table}[!ht]
\centering
\caption{\label{tab:mc_rmse_tau075}
Monte Carlo results (normal kernel, $\tau=0.75$): RMSE with scenarios S1--S4.}
\fontsize{9}{11}\selectfont
\begin{tabular}[t]{llcccccccc}
\toprule
 & & \multicolumn{4}{c}{S1} & \multicolumn{4}{c}{S2} \\
\cmidrule(l{3pt}r{3pt}){3-6} \cmidrule(l{3pt}r{3pt}){7-10}
 & Parameter & $n=75$ & $n=125$ & $n=200$ & $n=400$ & $n=75$ & $n=125$ & $n=200$ & $n=400$\\
\midrule
 & $\alpha$    & 0.3069 & 0.1838 & 0.1248 & 0.0884 & 0.0488 & 0.0281 & 0.0210 & 0.0125 \\
 & $\beta_{1}$ & 0.0446 & 0.0381 & 0.0267 & 0.0203 & 0.0893 & 0.0762 & 0.0534 & 0.0406 \\
 & $\beta_{2}$ & 0.0484 & 0.0335 & 0.0277 & 0.0196 & 0.0968 & 0.0670 & 0.0553 & 0.0392 \\
 & $\phi_{1}$  & 0.1360 & 0.0992 & 0.0766 & 0.0497 & 0.1360 & 0.0992 & 0.0766 & 0.0497 \\
 & $\phi_{2}$  & 0.1167 & 0.0884 & 0.0730 & 0.0495 & 0.1167 & 0.0884 & 0.0730 & 0.0495 \\
 & $\sigma$    & 0.0089 & 0.0071 & 0.0047 & 0.0038 & 0.0178 & 0.0142 & 0.0095 & 0.0076 \\
\midrule
 & & \multicolumn{4}{c}{S3} & \multicolumn{4}{c}{S4} \\
\cmidrule(l{3pt}r{3pt}){3-6} \cmidrule(l{3pt}r{3pt}){7-10}
 & Parameter & $n=75$ & $n=125$ & $n=200$ & $n=400$ & $n=75$ & $n=125$ & $n=200$ & $n=400$\\
\midrule
 & $\alpha$     & 0.2538 & 0.2076 & 0.1365 & 0.1088 & 0.4617 & 0.3609 & 0.2590 & 0.1570 \\
 & $\beta_{1}$  & 0.0941 & 0.0725 & 0.0576 & 0.0403 & 0.1457 & 0.1172 & 0.0925 & 0.0678 \\
 & $\beta_{2}$  & 0.0888 & 0.0730 & 0.0595 & 0.0418 & 0.1286 & 0.1063 & 0.0843 & 0.0633 \\
 & $\phi_{1}$   & 0.0964 & 0.0730 & 0.0434 & 0.0318 & 0.0902 & 0.0701 & 0.0486 & 0.0313 \\
 & $\sigma$     & 0.0958 & 0.0975 & 0.0979 & 0.0984 & 0.0204 & 0.0163 & 0.0131 & 0.0131 \\
 & $\theta_{1}$ & 0.1046 & 0.1024 & 0.1025 & 0.1016 & 0.0257 & 0.0192 & 0.0157 & 0.0145 \\
\bottomrule
\end{tabular}
\end{table}


\section{Application to proportion of stored hydroelectric energy data}
\label{Sec:application}

This section presents an empirical application of the proposed ULS--ARMA model to a
bounded hydrological time series dataset, the monthly proportion of stored
hydroelectric energy in the Southeast region of Brazil. The data consist of monthly observations
from May 2000 to August 2019, totaling 232 months, and are publicly available at \url{https://github.com/tatianefribeiro/ubxiiarma}; for more details, the interested reader can see \citep{Ribeiroeal2023} and Table~\ref{tab:stored_energy_data}. The Southeast subsystem plays a
central role in the Brazilian interconnected power system and, therefore, monitoring and
forecasting stored energy levels is crucial for planning and risk assessment, especially in
periods associated with water scarcity and supply stress. We compare the performance of the proposed models with alternative dynamic models for bounded data that
have been previously applied to this same series. In particular, we consider the
unit Burr XII quantile autoregressive moving average (UBXII--ARMA) model of
\citet{Ribeiroeal2023}, the beta autoregressive moving average (betaARMA) introduced by
\citet{RochaCribari2009} within the beta regression framework of
\citet{FerrariCribari2004}, and the Kumaraswamy autoregressive moving average (KARMA)
model proposed by \citet{Bayer2017}. These models represent well-established approaches
for modeling and forecasting proportion-valued time series and therefore serve as natural
benchmarks for assessing the advantages of the proposed ULS--ARMA specifications.

The series exhibits a pronounced annual seasonal pattern. In order to account for seasonality in a parsimonious way, we follow \citet{Ribeiroeal2023} and adopt harmonic terms as covariates, using the trigonometric regressors
$\cos(2\pi t/12)$ and $\sin(2\pi t/12)$, as in a simple harmonic regression representation.
In addition, in order to capture periods of persistently lower storage levels reported in the literature,
we include a crisis indicator (denoted by ${\rm crisis}_t$) as an exogenous covariate. Hence,
the competing models are fitted under a common regression structure with covariates
$(\cos(2\pi t/12),\,\sin(2\pi t/12),\,{\rm crisis}_t)$.

\subsection{Parameter estimation}
\label{subsec:params}

Table~\ref{tab:param_all} reports the conditional maximum likelihood estimates and standard
errors for all competing specifications, obtained by modeling the conditional median
(i.e., $\tau=0.5$), which goes in line with previous studies on this same series \citep{Ribeiroeal2023}. Overall, the results provide strong empirical support for dynamic modeling:
in all cases, the autoregressive parameters are highly significant and indicate pronounced
persistence in the conditional dynamics of the stored-energy proportion. Across the benchmark
specifications, namely UBXII--ARMA, betaAR, and KARMA, an AR(2) structure is selected, with
$\widehat{\phi}_1$ ranging from 1.32 to 1.61 and $\widehat{\phi}_2$ between $-0.41$ and
$-0.67$, reflecting a highly persistent behavior. A similar AR(2) structure
is also selected by the ULS--ARMA model with normal kernel, whose autoregressive
coefficients closely match those obtained from the existing unit-valued dynamic models.

In contrast, the ULS--ARMA specification based on the heavier-tailed Student-$t$ kernel leads to a
different and more flexible dynamic representation. The Student-$t$ ULS--ARMA model selects an
ARMA$(1,1)$ structure ($\widehat{\nu}=3$), combining a highly persistent first-order autoregressive component
with a statistically significant moving-average term. This indicates that, once robustness to
extreme observations is introduced through the Student-$t$ kernel, part of the serial
dependence is more effectively captured through short-run innovations rather than additional
autoregressive lags.

Regarding the regression component, the estimated covariate effects are remarkably stable
across all specifications. In particular, the coefficients associated with the seasonal harmonic
terms ($\beta_1$ and $\beta_2$) are positive and statistically significant at conventional levels
for all models, confirming the relevance of seasonal patterns in explaining the conditional
quantile dynamics of stored energy. In contrast, the crisis indicator ($\beta_3$) exhibits weaker
and less robust effects: it is marginally significant under the UBXII-ARMA model, but becomes
statistically insignificant under betaARMA, KARMA, and all ULS--ARMA variants. This finding
suggests that, once serial dependence and seasonality are properly accounted for, the remaining
impact of the crisis dummy on the conditional median is limited.

\begin{table}[!ht]
\small
\centering
\caption{Parameter estimates for the competing models.}
\begin{tabular}{llrrrr}
  \hline
Model & Parameter & Estimate & Std. Error & $z$ value & Pr($>|z|$) \\
  \hline
UBXII-ARMA & $\alpha$ & 0.0206 & 0.0156 & 1.3260 & 0.1848 \\
UBXII-ARMA & $\beta_1$ & 0.4034 & 0.0472 & 8.5449 & 0.0000 \\
UBXII-ARMA & $\beta_2$ & 0.1138 & 0.0419 & 2.7172 & 0.0066 \\
UBXII-ARMA & $\beta_3$ & -0.2630 & 0.1316 & -1.9988 & 0.0456 \\
UBXII-ARMA & $\phi_1$ & 1.3222 & 0.0432 & 30.5828 & 0.0000 \\
UBXII-ARMA & $\phi_2$ & -0.4072 & 0.0430 & -9.4752 & 0.0000 \\
UBXII-ARMA & $c$ & 11.3464 & 0.6468 & 17.5430 & 0.0000 \\
betaARMA & $\alpha$ & 0.0071 & 0.0097 & 0.7267 & 0.4674 \\
betaARMA & $\phi_1$ & 1.3797 & 0.0504 & 27.3683 & 0.0000 \\
betaARMA & $\phi_2$ & -0.4170 & 0.0506 & -8.2343 & 0.0000 \\
betaARMA & $\varphi$ & 200.7800 & 19.1045 & 10.5096 & 0.0000 \\
betaARMA & $\beta_1$ & 0.6172 & 0.0402 & 15.3585 & 0.0000 \\
betaARMA & $\beta_2$ & 0.1791 & 0.0395 & 4.5336 & 0.0000 \\
betaARMA & $\beta_3$ & 0.0155 & 0.0948 & 0.1632 & 0.8704 \\
KARMA & $\alpha$ & 0.0304 & 0.0132 & 2.3106 & 0.0209 \\
KARMA & $\phi_1$ & 1.6120 & 0.0644 & 25.0486 & 0.0000 \\
KARMA & $\phi_2$ & -0.6674 & 0.0621 & -10.7487 & 0.0000 \\
KARMA & $\varphi$ & 14.6954 & 0.7177 & 20.4769 & 0.0000 \\
KARMA & $\beta_1$ & 0.8756 & 0.0637 & 13.7389 & 0.0000 \\
KARMA & $\beta_2$ & 0.3578 & 0.0869 & 4.1184 & 0.0000 \\
KARMA & $\beta_3$ & 0.0912 & 0.0746 & 1.2232 & 0.2213 \\
ULS-ARMA (Normal) & $\alpha$ & 0.0073 & 0.0114 & 0.6348 & 0.5256 \\
ULS-ARMA (Normal) & $\beta_1$ & 0.6181 & 0.0462 & 13.3835 & 0.0000 \\
ULS-ARMA (Normal) & $\beta_2$ & 0.1910 & 0.0462 & 4.1333 & 0.0000 \\
ULS-ARMA (Normal) & $\beta_3$ & 0.0255 & 0.1105 & 0.2306 & 0.8176 \\
ULS-ARMA (Normal) & $\phi_1$ & 1.3823 & 0.0626 & 22.0645 & 0.0000 \\
ULS-ARMA (Normal) & $\phi_2$ & -0.4158 & 0.0622 & -6.6890 & 0.0000 \\
ULS-ARMA (Normal) & $\sigma$ & 0.1604 & 0.0076 & 20.9762 & 0.0000 \\
ULS-ARMA ($t$) & $\alpha$ & -0.0133 & 0.0138 & -0.9621 & 0.3360 \\
ULS-ARMA ($t$) & $\beta_1$ & 0.5535 & 0.0356 & 15.5379 & 0.0000 \\
ULS-ARMA ($t$) & $\beta_2$ & 0.1900 & 0.0340 & 5.5902 & 0.0000 \\
ULS-ARMA ($t$) & $\beta_3$ & 0.1406 & 0.1163 & 1.2086 & 0.2268 \\
ULS-ARMA ($t$) & $\phi_1$ & 0.9539 & 0.0148 & 64.5236 & 0.0000 \\
ULS-ARMA ($t$) & $\theta_1$ & 0.0591 & 0.0082 & 7.2280 & 0.0000 \\
ULS-ARMA ($t$) & $\sigma$ & 0.1076 & 0.0072 & 14.8435 & 0.0000 \\
  \hline
\end{tabular}
\label{tab:param_all}
\end{table}

In the work of \citet{Ribeiroeal2023}, the UBXII--ARMA model was shown to outperform both
the betaARMA model introduced by \citet{RochaCribari2009} and the KARMA model proposed by \citet{Bayer2017}. Motivated by these findings, we conduct a focused
comparison between the UBXII--ARMA model and the proposed ULS--ARMA specification with
Student-$t$ kernel across a fine grid of quantiles
$\tau \in \{0.01,0.02,\ldots,0.99\}$, evaluating model adequacy in terms of log-likelihood and
information criteria. In this sense, Table~\ref{tab:avg_ic_tau} summarizes the averages of the log-likelihood and information
criteria computed across the quantile grid. The results show that the ULS--ARMA model with
Student-$t$ kernel consistently dominates the UBXII--ARMA model, yielding a substantially
higher average log-likelihood and uniformly lower AIC, BIC, CAIC, and HQIC values. These
findings indicate that the gains achieved by the proposed model are not restricted to a specific
quantile level, but rather persist across the entire conditional distribution, highlighting the
benefits of combining quantile-based dynamics with a flexible unit-log-symmetric distribution
capable of accommodating heavier tails.

\begin{table}[!ht]
\centering
\caption{Averages of the log-likelihood and information criteria computed across
$\tau \in \{0.01,0.02,\ldots,0.99\}$ for the competing quantile-based models.}
\label{tab:avg_ic_tau}
\renewcommand{\arraystretch}{1.15}
\setlength{\tabcolsep}{10pt}
\begin{tabular}{lcc}
\toprule
Indicator & UBXII--ARMA & ULS--ARMA (Student-$t$) \\
\midrule
log-likelihood & $410.168$ & \textbf{$444.331$} \\
AIC  & $-806.013$ & \textbf{$-873.953$} \\
BIC  & $-781.645$ & \textbf{$-848.927$} \\
CAIC & $-806.013$ & \textbf{$-873.953$} \\
HQIC & $-808.256$ & \textbf{$-876.256$} \\
\bottomrule
\end{tabular}
\end{table}


\subsection{Out-of-sample forecasting performance}
\label{subsec:forecasting}

In order to evaluate predictive accuracy, we computed out-of-sample forecast errors for horizons $h=1,\ldots,10$ using the mean squared error (MSE) and mean absolute percentage error (MAPE). Results are summarized in Tab.~\ref{tab:forecast_horizons}. For very short horizons ($h=1$ and $h=2$), KARMA(2) achieves the smallest MSE and MAPE; however, its performance deteriorates drastically for $h\ge 3$, producing extremely large errors relative to all other models. This indicates that, although KARMA(2) may fit local one-step variation, it does not provide stable multi-step dynamics for this dataset. In contrast, the ULS-ARMA model with Student-$t$ kernel delivers the most consistent multi-step forecasting performance. From $h=3$ onward, the Student-$t$ ULS-ARMA attains the lowest MSE for all horizons $h=3,\ldots,10$, and simultaneously yields the smallest MAPE in the same range (Tab.~\ref{tab:forecast_horizons}). In practical terms, this means that the ULS-ARMA specification provides superior medium- and long-horizon predictions for the stored-energy proportion, which is particularly relevant for operational planning contexts where decision-making often relies on forecasts beyond one or two periods ahead.

The UBXII-AR(2), betaARMA(2), and the Normal ULS-ARMA alternative show intermediate performance: they remain stable across horizons but are consistently dominated by the Student-$t$ ULS-ARMA for $h\ge 3$. This suggests that allowing heavier tails via the Student-$t$ kernel is beneficial for capturing occasional atypical movements in the stored-energy proportion, improving the robustness of the conditional quantile recursion and reducing forecast error accumulation.

\begin{table}[!ht]
\small
\centering
\caption{Out-of-sample forecast errors for prediction horizons $h = 1,\ldots,10$.}
\label{tab:forecast_horizons}
\renewcommand{\arraystretch}{1.15}
\setlength{\tabcolsep}{2pt}

\begin{tabular}{lcccccccccc}
\toprule
 & \multicolumn{10}{c}{Forecast horizon $h$} \\
\cmidrule(lr){2-11}
Model & $h=1$ & $h=2$ & $h=3$ & $h=4$ & $h=5$ & $h=6$ & $h=7$ & $h=8$ & $h=9$ & $h=10$ \\
\midrule
\multicolumn{11}{l}{\textbf{(a) Mean Squared Error (MSE)}} \\
\midrule
UBXII-ARMA(2)        & 0.0008 & 0.0008 & 0.0010 & 0.0026 & 0.0025 & 0.0024 & 0.0022 & 0.0020 & 0.0018 & 0.0017 \\
betaARMA(2)          & 0.0011 & 0.0010 & 0.0013 & 0.0038 & 0.0045 & 0.0050 & 0.0053 & 0.0049 & 0.0044 & 0.0039 \\
KAR(2)             & \textbf{0.0005} & \textbf{0.0003} & 0.0063 & 0.0221 & 0.0345 & 0.0456 & 0.0540 & 0.0586 & 0.0598 & 0.0586 \\
ULS-ARMA (Normal)  & 0.0011 & 0.0011 & 0.0013 & 0.0036 & 0.0042 & 0.0047 & 0.0049 & 0.0046 & 0.0041 & 0.0037 \\
ULS-ARMA (t)       & 0.0014 & 0.0017 & \textbf{0.0012} & \textbf{0.0018} & \textbf{0.0016} & \textbf{0.0014} & \textbf{0.0012} & \textbf{0.0011} & \textbf{0.0012} & \textbf{0.0015} \\
\midrule
\multicolumn{11}{l}{\textbf{(b) Mean Absolute Percentage Error (MAPE)}} \\
\midrule
UBXII-ARMA(2)        & 11.88 & 10.76 & 12.04 & 16.23 & 15.28 & 14.29 & 13.45 & 12.44 & 11.43 & 10.96 \\
betaARMA(2)          & 13.52 & 12.20 & 13.68 & 19.34 & 19.69 & 19.67 & 19.31 & 18.22 & 16.51 & 14.97 \\
KAR(2)             & \textbf{9.46} & \textbf{6.11} & 20.88 & 38.08 & 44.97 & 49.22 & 51.99 & 53.46 & 54.01 & 54.16 \\
ULS-ARMA (Normal)  & 13.91 & 12.80 & 13.65 & 18.97 & 19.19 & 19.15 & 18.81 & 17.75 & 16.06 & 14.61 \\
ULS-ARMA (t)       & 15.56 & 15.82 & \textbf{12.52} & \textbf{14.61} & \textbf{12.96} & \textbf{11.64} & \textbf{10.39} & \textbf{9.50} & \textbf{9.62} & \textbf{10.13} \\
\bottomrule
\end{tabular}
\end{table}

In summary, the application confirms that dynamic unit-log-symmetric modeling is well-suited for bounded energy storage proportions. The fitted models capture strong serial persistence and meaningful covariate effects in the conditional quantile. Among all competitors, the Student-$t$ ULS-ARMA specification stands out as the most reliable approach for multi-step forecasting, achieving the best overall out-of-sample performance for horizons $h=3$ to $h=10$ (Tab.~\ref{tab:forecast_horizons}). These results support the usefulness of combining quantile-based dynamics with a flexible unit-log-symmetric distributional structure.

\subsection{Residual analysis}
\label{subsec:residuals}

In order to evaluate the adequacy of the fitted ULS--ARMA models, we carry out a
residual analysis based on two complementary diagnostics: the generalized Cox--Snell
residuals and the randomized quantile residuals. These tools allow us to assess whether
the assumed conditional distributions are consistent with the observed data, and whether
any systematic departure from the fitted model remains unexplained.

Let $\widehat{q}_{\tau,t}$ denote the fitted conditional $\tau$-quantile at time $t$ and
let $\widehat{\sigma}$ denote the estimated scale parameter. The generalized Cox--Snell
(GCS) residual is defined as
\begin{equation}\label{eq:cs_resid}
r_{{\rm CS},t}
=
-\log\!\Big(1 - F_t(y_t)\Big),
\end{equation}
where $F_t(y_t) = F_{Y_t \mid \mathcal{A}_{t-1}}(y_t;\,\widehat{q}_{\tau,t},\widehat{\sigma})$
is the fitted conditional CDF of $Y_t$ evaluated at the observed value $y_t$.
Under a correctly specified model, $r_{{\rm CS},t}$ follows approximately an
$\mathrm{Exp}(1)$ distribution, so that a QQ plot of the GCS residuals against
$\mathrm{Exp}(1)$ quantiles provides a visual check of distributional adequacy.

The randomized quantile (RQ) residual is defined as
\begin{equation}\label{eq:rq_resid}
r_{{\rm q},t}
=
\Phi^{-1}\!\Big(F_t(y_t)\Big),
\end{equation}
where $\Phi^{-1}(\cdot)$ denotes the standard normal quantile function. Under a correctly
specified model, $r_{{\rm q},t}$ follows a standard normal distribution, whereas systematic
departures from normality in a QQ plot of $r_{{\rm q},t}$ against $\mathrm{N}(0,1)$
quantiles indicate model misspecification.

\begin{figure}[!ht]
\centering
\subfigure[GCS residuals, ULS-ARMA (Normal)]
  {\includegraphics[width=0.40\textwidth]{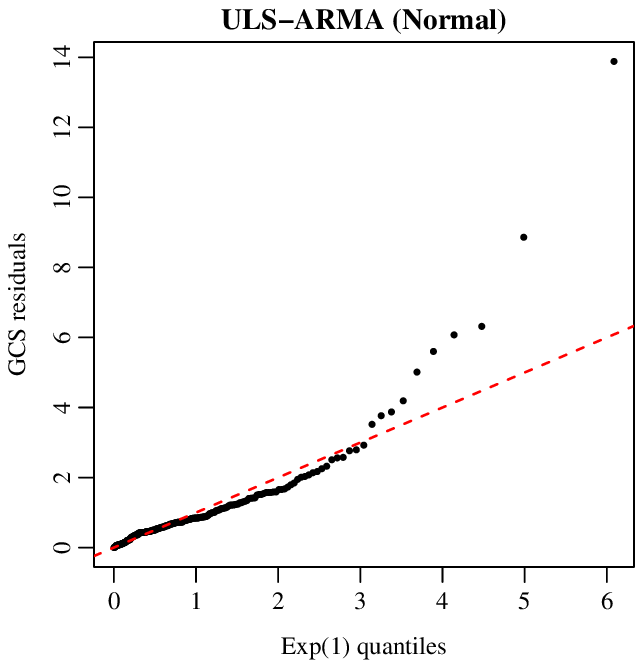}}
\hfill
\subfigure[RQ residuals, ULS-ARMA (Normal)]
  {\includegraphics[width=0.40\textwidth]{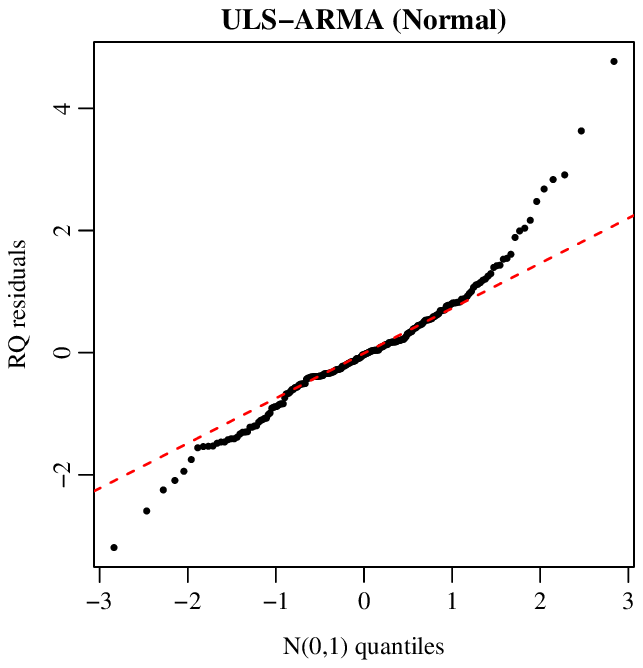}}
\\[4pt]
\subfigure[GCS residuals, ULS-ARMA ($t$)]
  {\includegraphics[width=0.40\textwidth]{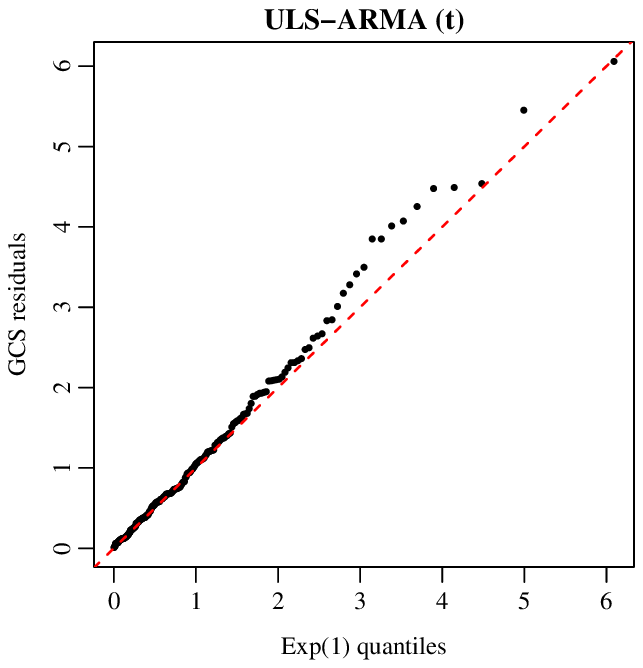}}
\hfill
\subfigure[RQ residuals, ULS-ARMA ($t$)]
  {\includegraphics[width=0.45\textwidth]{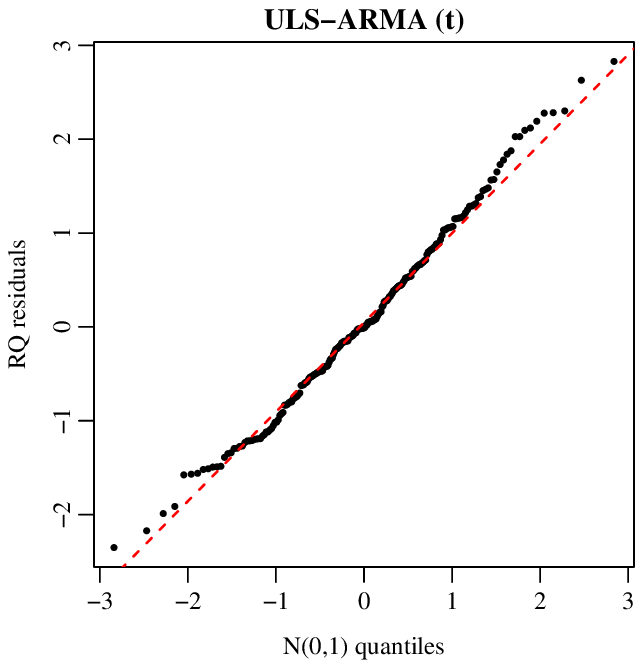}}
\caption{QQ plots of the generalized Cox--Snell residuals (left panels) and the
randomized quantile residuals (right panels) for the ULS--ARMA model with normal kernel
(AR(2), top row) and Student-$t$ kernel (ARMA$(1,1)$, $\hat{\nu}=3$, bottom row).}
\label{fig:qqplots_resid}
\end{figure}

Figure~\ref{fig:qqplots_resid} displays the QQ plots of both residual types for the two
ULS--ARMA specifications, namely the model with normal kernel (AR(2) structure) and the
model with Student-$t$ kernel (ARMA$(1,1)$, $\widehat{\nu}=3$). For the GCS residuals
(panels (a) and (c)), the points align closely with the $\mathrm{Exp}(1)$ reference line
in both cases, indicating that the fitted conditional distributions provide an adequate
description of the observed proportions. For the RQ residuals (panels (b) and (d)), the
Student-$t$ specification exhibits a slightly tighter adherence to the $\mathrm{N}(0,1)$
reference line, in particular in the tails, which is consistent with the superior
predictive performance documented in Section~\ref{subsec:forecasting}. Overall, the
residual analysis supports the adequacy of both ULS--ARMA specifications and confirms that
the conditional distribution is well-calibrated, with a marginal advantage for the heavy-tailed
formulation.

\section{Concluding remarks} \label{Sec:concluding}
\noindent

It is widely known that bounded time series data, such as proportions and rates, arise frequently in environmental, economic, and hydrological applications and require dedicated distributional and inferential frameworks. In this paper, we proposed the quantile unit-log-symmetric autoregressive moving average (QULS--ARMA) model for bounded time series. The model extends the unit-log-symmetric distribution to a dynamic setting by introducing ARMA-type dependence directly in the conditional quantile, allowing for flexible distributional shapes through the normal and Student-$t$ log-symmetric kernels. This construction provides a framework for proportion data arising from ratios of dependent positive components, whereas serial dependence and asymmetric dynamics are explicitly accommodated through the ARMA quantile structure. Theoretical formulation and conditional maximum likelihood estimation were developed under a quantile parameterization, and Monte Carlo simulation studies showed good finite-sample performance of the proposed estimators across a range of scenarios, quantile levels $\tau\in\{0.25,0.50,0.75\}$, and kernel specifications. The residual analysis based on generalized Cox--Snell and randomized quantile residuals confirmed the adequacy of the fitted conditional distributions. An empirical application to monthly proportions of stored hydroelectric energy in Brazil illustrated the practical relevance of the proposed model. The proposed ULS--ARMA specification, especially under the Student-$t$ kernel, provided superior multi-step forecasting performance when compared with established competitors. Extensions to multivariate bounded time series and seasonal autoregressive moving average models are currently under investigation.


\paragraph{Acknowledgments}
This study was financed in part by the Coordenação de 
Aperfeiçoamento de Pessoal de Nível Superior - Brasil (CAPES) 
(Finance Code 001). Roberto Vila and Helton Saulo gratefully acknowledge financial support from FAP-DF and CNPq, Brazil.

\paragraph{Disclosure statement}
There are no conflicts of interest to disclose.




\appendix
\begin{table}[!ht]
\centering
\caption{Monthly proportion of stored hydroelectric energy in the Southeast region of Brazil.}
\label{tab:stored_energy_data}
\renewcommand{\arraystretch}{1.05}
\resizebox{\textwidth}{!}{%
\begin{tabular}{c cccccccccccc}
\toprule
Year & Jan & Feb & Mar & Apr & May & Jun & Jul & Aug & Sep & Oct & Nov & Dec \\
\midrule
2000 &  &  &  &  & 0.536800 & 0.471400 & 0.399200 & 0.322700 & 0.308400 & 0.230500 & 0.222100 & 0.286700 \\
2001 & 0.315600 & 0.335900 & 0.346800 & 0.323000 & 0.298800 & 0.287100 & 0.268500 & 0.234600 & 0.206900 & 0.211300 & 0.231900 & 0.325300 \\
2002 & 0.470600 & 0.633300 & 0.703700 & 0.692300 & 0.674700 & 0.648200 & 0.596000 & 0.536200 & 0.491500 & 0.409600 & 0.384400 & 0.414100 \\
2003 & 0.588000 & 0.679600 & 0.744100 & 0.776200 & 0.761400 & 0.728800 & 0.673200 & 0.591200 & 0.501000 & 0.407800 & 0.360000 & 0.373400 \\
2004 & 0.473600 & 0.664600 & 0.759200 & 0.807800 & 0.826800 & 0.822500 & 0.803300 & 0.745000 & 0.659200 & 0.617900 & 0.593000 & 0.645400 \\
2005 & 0.757600 & 0.787400 & 0.859100 & 0.857400 & 0.853400 & 0.825600 & 0.781600 & 0.700600 & 0.652700 & 0.603500 & 0.592200 & 0.671000 \\
2006 & 0.711400 & 0.784600 & 0.853300 & 0.873200 & 0.846700 & 0.783000 & 0.701300 & 0.593800 & 0.503400 & 0.456300 & 0.428000 & 0.536600 \\
2007 & 0.785500 & 0.846100 & 0.873300 & 0.878100 & 0.867700 & 0.831800 & 0.795200 & 0.720100 & 0.620300 & 0.516100 & 0.481800 & 0.461200 \\
2008 & 0.507700 & 0.655800 & 0.785000 & 0.822400 & 0.829000 & 0.795500 & 0.731500 & 0.664000 & 0.578500 & 0.519100 & 0.497300 & 0.558200 \\
2009 & 0.661100 & 0.760700 & 0.805800 & 0.835500 & 0.821900 & 0.785700 & 0.761000 & 0.724800 & 0.702200 & 0.691200 & 0.673100 & 0.723000 \\
2010 & 0.766900 & 0.778400 & 0.825100 & 0.818100 & 0.786400 & 0.743500 & 0.678800 & 0.585500 & 0.492500 & 0.431000 & 0.406400 & 0.447700 \\
2011 & 0.629900 & 0.681200 & 0.829200 & 0.878000 & 0.878200 & 0.852700 & 0.806200 & 0.740700 & 0.653400 & 0.613700 & 0.570300 & 0.606400 \\
2012 & 0.760700 & 0.800100 & 0.784800 & 0.760400 & 0.724100 & 0.724300 & 0.668900 & 0.575100 & 0.478900 & 0.370200 & 0.318900 & 0.288600 \\
2013 & 0.374700 & 0.456000 & 0.541600 & 0.625500 & 0.627500 & 0.637700 & 0.608600 & 0.551300 & 0.486900 & 0.450700 & 0.416300 & 0.430100 \\
2014 & 0.401700 & 0.347300 & 0.363500 & 0.380300 & 0.373400 & 0.364300 & 0.343200 & 0.303100 & 0.251700 & 0.187400 & 0.158200 & 0.193100 \\
2015 & 0.170400 & 0.206900 & 0.285400 & 0.335800 & 0.360600 & 0.361500 & 0.374200 & 0.342900 & 0.322800 & 0.277400 & 0.275000 & 0.297900 \\
2016 & 0.444300 & 0.509300 & 0.582800 & 0.577100 & 0.568500 & 0.561400 & 0.515900 & 0.459400 & 0.401600 & 0.348300 & 0.334700 & 0.337400 \\
2017 & 0.373300 & 0.403800 & 0.414500 & 0.418600 & 0.433200 & 0.421200 & 0.381600 & 0.325000 & 0.242100 & 0.176100 & 0.187400 & 0.226000 \\
2018 & 0.311333 & 0.368382 & 0.423658 & 0.439873 & 0.425606 & 0.397108 & 0.342491 & 0.279828 & 0.229317 & 0.202239 &  &  \\
\bottomrule
\end{tabular}}
\end{table}

\end{document}